# A surge of light at the birth of a supernova

Bersten, M. C.[1;2;3], Folatelli, G.[1;2;3], García, F.[2;4;5], Van Dyk, S. D.[6], Benvenuto, O. G.[1;2], Orellana, M.[7], Buso, V.[8], Sánchez, J. L.[9], Tanaka, M.[10], Maeda, K.[3;11], Filippenko, A. V.[12;13], Zheng, W.,[12], Brink, T. G.[12], Cenko, S. B.[14;15], De Jaeger, T.[12], Kumar, S.[16], Moriya, T. J.[10], Nomoto, K.[3], Perley, D. A.[17], Shivvers, I.[12] & Smith, N.[18]

[1]Instituto de Astrofísica de La Plata (IALP), CONICET, Argentina.
[2]Facultad de Ciencias Astronómicas y Geofísicas, Universidad Nacional de La Plata, Paseo del Bosque s/n, B1900FWA La Plata, Argentina.
[3]Kavli Institute for the Physics and Mathematics of the Universe, Todai Institutes for Advanced Study, University of Tokyo, 5-1-5 Kashiwanoha, Kashiwa, Chiba 277-8583, Japan.
[4]Instituto Argentino de Radioastronomía (CCT-La Plata, CONICET; CICPBA), C.C. No. 5, 1894,Villa Elisa, Argentina.
[5]Paris Diderot, AIM, Sorbonne Paris Cité, CEA, CNRS, F-91191 Gif-sur-Yvette, France.
[6]Caltech/IPAC, Mailcode 100-22, Pasadena, CA 91125, USA.
[6]Member of the Carrera del Investigador Científico de la Comisión de Investigaciones Científicas de la Provincia de Buenos Aires (CIC), Argentina.
[7]Sede Andina, Universidad Nacional de Río Negro, Mitre 630 (8400) Bariloche, CONICET, Argentina.
[8]Observatorio Astronómico Busoniano, Entre Ríos 2974 (2000), Rosario, Argentina.
[9]Observatorio Astronómico Geminis Austral, Rosario, Argentina.
[10]Division of Theoretical Astronomy, National Astronomical Observatory of Japan, National Institutes of Natural Sciences, 2-21-1 Osawa, Mitaka, Tokyo 181-8588, Japan.
[11]Department of Astronomy, Kyoto University, Kitashirakawa-Oiwake-cho, Sakyo-ku, Kyoto 606-8502, Japan.
[12]Department of Astronomy, University of California, Berkeley, CA 94720-3411, USA.
[13]Miller Senior Fellow, Miller Institute for Basic Research in Science, University of California, Berkeley, CA 94720, USA.
[14]Astrophysics Science Division, NASA Goddard Space Flight Center, Greenbelt, MD 20771, USA.
[15]Joint Space-Science Institute, University of Maryland, College Park, MD 20742, USA.
[16]Department of Physics, Florida State University, 77 Chieftain Way, Tallahassee, Florida 32306, USA.
[17]Astrophysics Research Institute, Liverpool John Moores University, IC2, Liverpool Science Park, 146 Brownlow Hill, Liverpool L3 5RF, UK.
[18]Steward Observatory, University of Arizona, 933 N. Cherry Ave., Tucson, AZ 85721, USA.

**An important problem in astrophysics is to elucidate the properties of massive stars that explode as supernovae [1, 2]. The electromagnetic emission during the first minutes to hours after the emergence of the shock from the stellar surface conveys unique information about the final evolution and structure of the exploding star [3, 4, 5, 6]. However, the unpredictable nature of the supernova event hinders the detection of this brief initial phase [7, 8, 9]. Here we report the serendipitous discovery of a newly-born normal Type IIb supernova [SN 2016gkg; 10] by amateur astronomer Víctor Buso, revealing an unprecedented optical rise rate of about 40 magnitudes per day. The very frequent sampling of the discovery observations allowed us to study in detail the outermost progenitor structure and the physics of the shock emergence. Our hydrodynamical models of the supernova naturally account for the complete supernova evolution over distinct phases that are regulated by different physical processes. This result suggests that it is appropriate to decouple the treatment of the shock propagation from the unknown mechanism that triggers the explosion.**

On September 20, 2016 (dates are given in UT throughout this paper), amateur astronomer Víctor Buso from Rosario, Argentina, set out to test a new camera mounted on his 40-cm Newtonian telescope. He decided to point at NGC 613, a spiral galaxy at a distance of 26.4 Mpc, because it was at that time located near the zenith. Over approximately one and a half hours he kept imaging the galaxy with a clear filter while a supernova (SN) was being born. Buso used 20 s exposures to avoid saturation caused by the bright city sky. An initial series of forty images obtained during 20 min showed noسign of the SN. From the combined image we obtained a 5-sigma detection limit of 19.4 mag converted to the V band (see Methods). When observations were resumed, after an interval of 45 min, the SN became visible. During the remaining 25 min of observations the SN doubled its flux (see Figure 1 and more details in Methods). The deep detection limit and frequent sampling during the initial rise constitute an unprecedented set of observations for an SN discovery. A linear fit to the discovery photometry yields a remarkably fast rise rate of 43±6 mag d$^{-1}$. An extrapolation of this linear rise back in time down to the proposed progenitor brightness of V ≈ 24 mag (see Methods) suggests that SN 2016gkg exploded some time between UT 2:50 and 5:35. This constraint of less than 3 hours on the explosion epoch is one of the most stringent available to date [cf. 11, 12, 13, 14, 15].

The prompt discovery and announcement of SN 2016gkg [10] triggered extensive monitoring that began less than one day later, including Swift X-ray, ultraviolet (UV), and optical observations [16, 17]. This permitted excellent coverage of the subsequent SN evolution. Consequently, the cooling peak, which lasted for about three days, is one of the best observed to date. Followup spectroscopy provided the classification as a Type IIb SN [16]. Our own photometric and spectroscopic monitoring starting less than one day after discovery is described in Methods. In addition to these observations, the Hubble Space Telescope (HST) archive contained images of the SN site obtained with the WFPC2 camera in 2001. An object was identified at the SN location in all three available optical HST bands (F450W, F606W, and F814W). Our analysis of these data and inferences about the progenitor system are given in Methods [see also 16, 18].

To interpret the physical process that governed the behaviour of the supernova at the time of discovery, we plot the luminosity versus the rise rate and compare these data with available early optical observations of other supernovae and transients (Fig. 2; see ref. 9 for a similar graph in the near-ultraviolet range). The low luminosity and fast rise seen in SN 2016gkg place the discovery data in a clearly different location on the diagram compared with later observations and with data for other objects. This suggests a different physical origin for the initial rise of SN 2016gkg.

A natural explanation for the above comparison is that the first signal of SN 2016gkg corresponds to the long sought-after shock breakout (SBO) phase [19, 20, 21]. Indeed, hydrodynamical simulations indicate that, although the SBO signal is predominant in the X-ray and UV ranges, it has a clear manifestation in the optical range as well, characterised by an extremely rapid brightening at relatively low luminosity [6, 22]. In order to test this interpretation, we performed numerical modelling of the SN observations.

We divided the SN modelling into two main stages, following [23]. First, the overall SN evolution was modelled by adjusting the explosion energy, ejecta mass, and $^{56}$Ni mass to reproduce the main light-curve peak and expansion velocities from spectral lines. In this way we found that a model with an energy of E ≈ 1.2 × 10$^{51}$ erg, an ejecta mass of $M_{ej}$ ≈ 3.4 $M_\odot$, and a $^{56}$Ni mass of $M_{Ni}$ ≈ 0.085 $M_\odot$ provided a good match to the observations (see Methods for alternative solutions). Once these parameters were constrained, the second step consisted in modelling the post-shock cooling peak. It is well known that in order to reproduce such a peak, an extended H-rich envelope has to be attached to the usual progenitor structure from stellar-evolution calculations. We varied the radius and mass of the envelope and arrived at a radius of $R_{env} \approx 320$ $R_\odot$ and a mass of $M_{env} \approx 0.01$ $M_\odot$ for the extended envelope [see 16, 17, for other estimates]. Our preferred model is shown in Figure 3 (see more details in Methods). The derived parameters are in close agreement with those of normal Type IIb SNe. In particular, the SN 2016gkg progenitor was slightly more massive and had a somewhat more extended envelope than the well-studied Type IIb SN 2011dh [23].

We note that, although the model was designed to match the observed cooling and radioactive peaks, it also explains without any modification the quick rise of the discovery data as being caused by the SBO emission. In fact, regardless of the adopted parameters, no physical process other than the SBO can produce such a fast rise. We verified this by exploring a set of hydrostatic progenitor structures and explosion parameters and comparing the rise slopes of our light-curve models during the SBO and the post-shock cooling phases (see Methods). We identified the explosion energy as the dominant factor that determines the slope of the rise to the cooling peak. However, even with an energy value far beyond what is allowed by the rest of the observations, we were unable to reach a rise rate of the cooling peak near that of the SBO phase. This implies that different processes govern the initial rise and the cooling phase, which provides strong support to our interpretation of the early rise as the manifestation of the SBO.

A closer look at the discovery data reveals that the observed rise is more gradual than that of the model. This difference could be caused by limitations of our radiative-transfer treatment[24,25], but it could also be indicative of the presence of some circumstellar material (Methods). More detailed analysis of the shock-breakout signal could potentially provide important information about the outermost progenitor structure and the physical processes that occur during the emergence of the shock. The serendipitous nature of the discovery observations and the sampling that was required highlight the difficulty of systematizing this type of finding, which has been the goal of several recent and future transient surveys, including KISS[7], HiTS[8], HSC-SHOOT[9], KEGS (http://www.mso.anu.edu.au/kegs/) and ZTF[26]. We note that the chance probability of this discovery is of the order of $10^{-6}$ assuming a duration of 1 h and one supernova per century per galaxy. If we consider other factors such as the sky conditions of the observing site and the location of the supernova away from bright host-galaxy regions, then this probability decreases by one order of magnitude.

**Supplementary information** is available in the online version of the paper.

**Acknowledgements** We are grateful to Peter Brown for providing information about the photometry of the early Swift/UVOT data of SN 2006aj. M.B. acknowledges support from the Agencia Nacional de Promoción Científica y Tecnológica (ANPCyT) through grant

PICT-2015-3083 "Progenitores de Supernovas de Colapso Gravitatorio," and from the Munich Institute for Astro- and Particle Physics (MIAPP) of the DFG cluster of excellence "Origin and Structure of the Universe." M.B., G.F., and O.G.B. acknowledge support from grant PIP-2015- 2017-11220150100746CO of CONICET "Estrellas Binarias y Supernovas." G.F. further acknowledges support from ANPCyT grant PICT-2015-2734 "Nacimiento y Muerte de Estrellas Masivas: Su relación con el Medio Interestelar." K.M. acknowledges support from JSPS KAKENHI Grant 17H02864. Partial support for this work was provided by NASA through programs GO-14115 and AR-14295 from the Space Telescope Science Institute, which is operated by AURA, Inc., under NASA contract NAS 5-26555. M.O. acknowledges support from grant PI UNRN40B531. A.V.F. is also grateful for financial assistance from the Christopher R. Redlich Fund, the TABASGO Foundation, and the Miller Institute for Basic Research in Science (U.C. Berkeley). We thank U.C. Berkeley undergraduate students S. Channa, G. Halevy, A. Halle, M. de Kouchkovsky, J. Molloy, T. Ross, S. Stegman, and S. Yunus for their effort in taking Lick/Nickel data. The Lick and Keck Observatory staff provided excellent assistance. A major upgrade of the Kast spectrograph on the Shane 3 m telescope at Lick Observatory was made possible through generous gifts from William and Marina Kast as well as the Heising-Simons Foundation. Research at Lick Observatory is partially supported by a generous gift from Google. KAIT and its ongoing operation were made possible by donations from Sun Microsystems, Inc., the Hewlett-Packard Company, AutoScope Corporation, Lick Observatory, the NSF, the University of California, the Sylvia & Jim Katzman Foundation, and the TABASGO Foundation. Some of the data presented herein were obtained at the W. M. Keck Observatory, which is operated as a scientific partnership among the California Institute of Technology, the University of California, and NASA; the observatory was made possible by the generous financial support of the W. M. Keck Foundation. O.G.B. is a member of the Carrera del Investigador Científico de la Comisión de Investigaciones Científicas de la Provincia de Buenos Aires (CIC), Argentina.

**Author contributions** M.B. hydrodynamical models, interpretation. G.F. SN and pre-SN data analysis, interpretation. F.G. SN data analysis, interpretation. S.V.D. SN and pre-SN data analysis, interpretation. O.G.B. binary evolution models. M.O. early data comparisons. M.T., K.M. SBO interpretation. V.B. SN discovery. J.L.S. early SN observations. A.V.F. Lick and Keck Observatory data, paper editing. W.Z., T.G.B., T.D.J, I.S., S.K and N.S. observations and reductions. T.M. CSM interpretation. K.N. pre-SN models. S.B.C., D.A.P. spectral reductions.

**Author information** Reprints and permissions information is available at www.nature.com/reprints. The authors declare that they have no competing financial interests. Correspondence and requests for materials should be addressed to M.B. (email: mbersten@fcaglp.unlp.edu.ar) and G.F. (email: gaston@fcaglp.unlp.edu.ar).

**Figure 1**

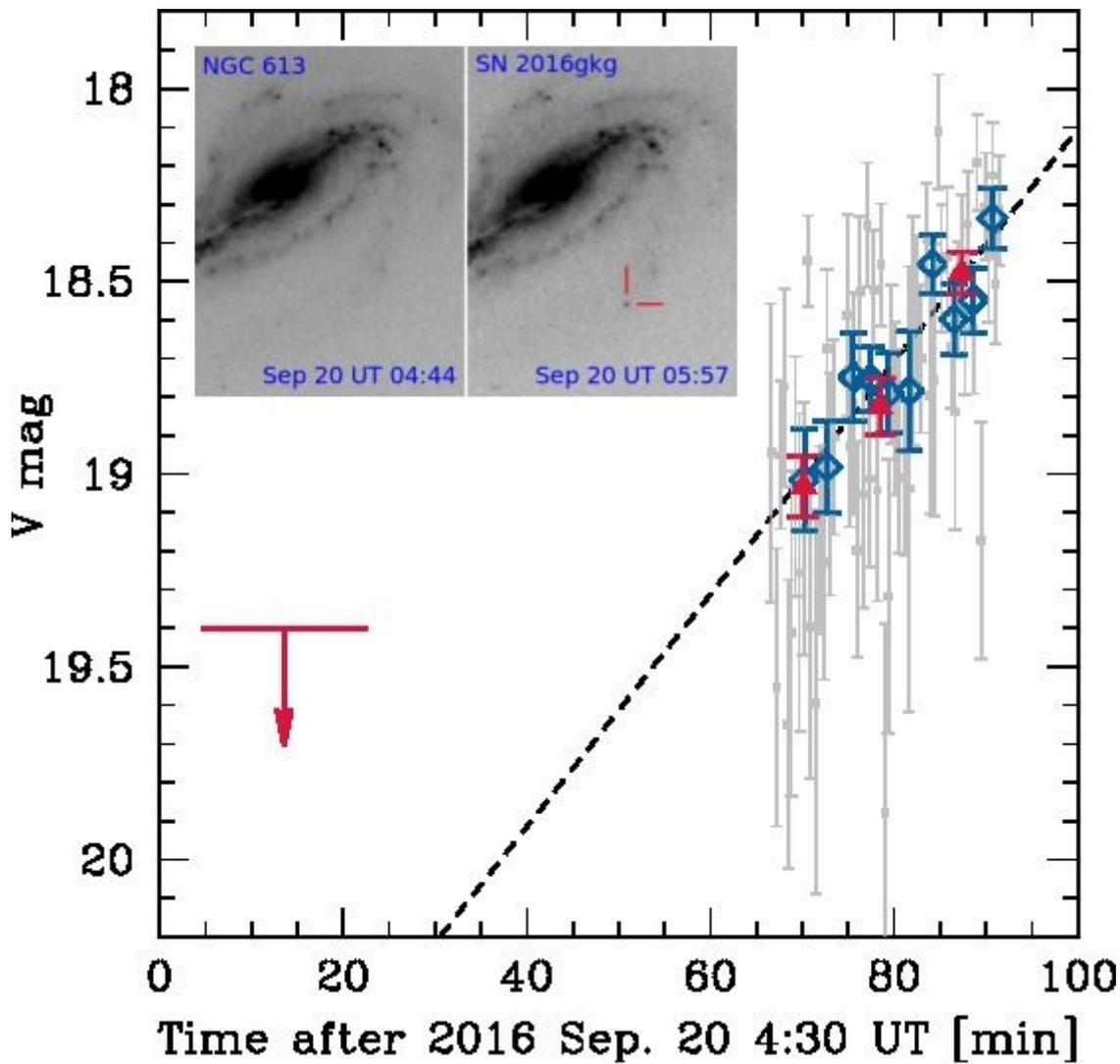

**Photometry of SN 2016gkg at discovery.** The data show a 5-sigma detection limit (red arrow) and sharp rise (points) starting less than 1 hr later. The inset images display a combination of the first series of 40 images (left), and a combination of the last series of 21 images (right). Photometry is shown for individual images (grey points), combinations of 5 or 6 images (blue diamonds), and combinations of 17-21 images (red triangles). The dashed line is a linear fit to the blue diamonds, with a slope of 43 ± 6 mag d$^{-1}$. Uncertainties are given as 1σ standard deviations. Photometry from combined images reveals hints of structure around the linear fit, although its statistical significance is low (the reduced χ$^2$ of the linear fit is 0.85).

**Figure 2**

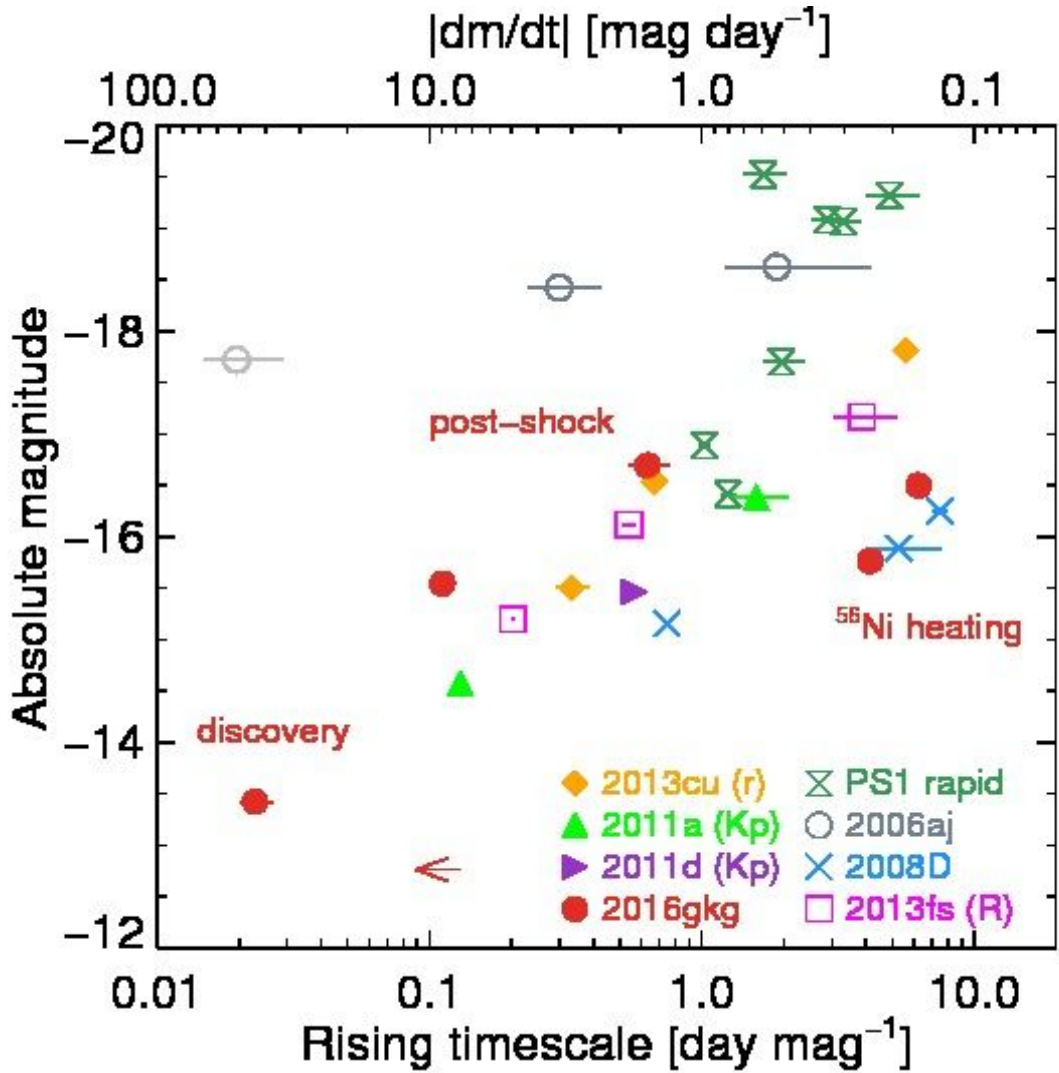

**Luminosity versus rise rate for objects with early optical detections.** The rate of change of magnitudes per unit time, d$m$/d$t$, is shown in the upper x axis. V-band data are used, except as indicated in parentheses next to the object name. SN 2016gkg is shown with red dots following its light-curve evolution, and an upper limit (arrow) from the prediscovery detection limit. The point labelled "discovery" represents the initial rise sampled by Víctor Buso. "PS1 rapid" indicates rapidly evolving transients from Pan-STARRS1 [29]. Each point is one of the following objects (in parentheses we give the observed band that best matches the V band, according to the redshift of each object): PS1-10ah (rP1), PS1-10bjp (rP1), PS1-11qr (iP1), PS1-11bbq (zP1), PS1-12bv (iP1), PS1-12brf (iP1), and PS1-13duy (iP1). Other SNe are as follows: the broad-lined Type Ic, GRB-associated SN 2006aj, with data from [30] (dark-grey circles) and [11] (light-grey circle; but see a discussion about this data point in Methods); the X-ray flash associated with the Type Ib SN 2008D, with data from [12]; the Type II SN 2013fs (iPTF13dqy) from [15]; the Type IIb SN 2013cu (iPTF13ast) from [13]; and the Type II SNe from the Kepler mission, KSN 2011a, and KSN 2011d [14]. Uncertainties are given as 1σ standard deviations.

**Figure 3**

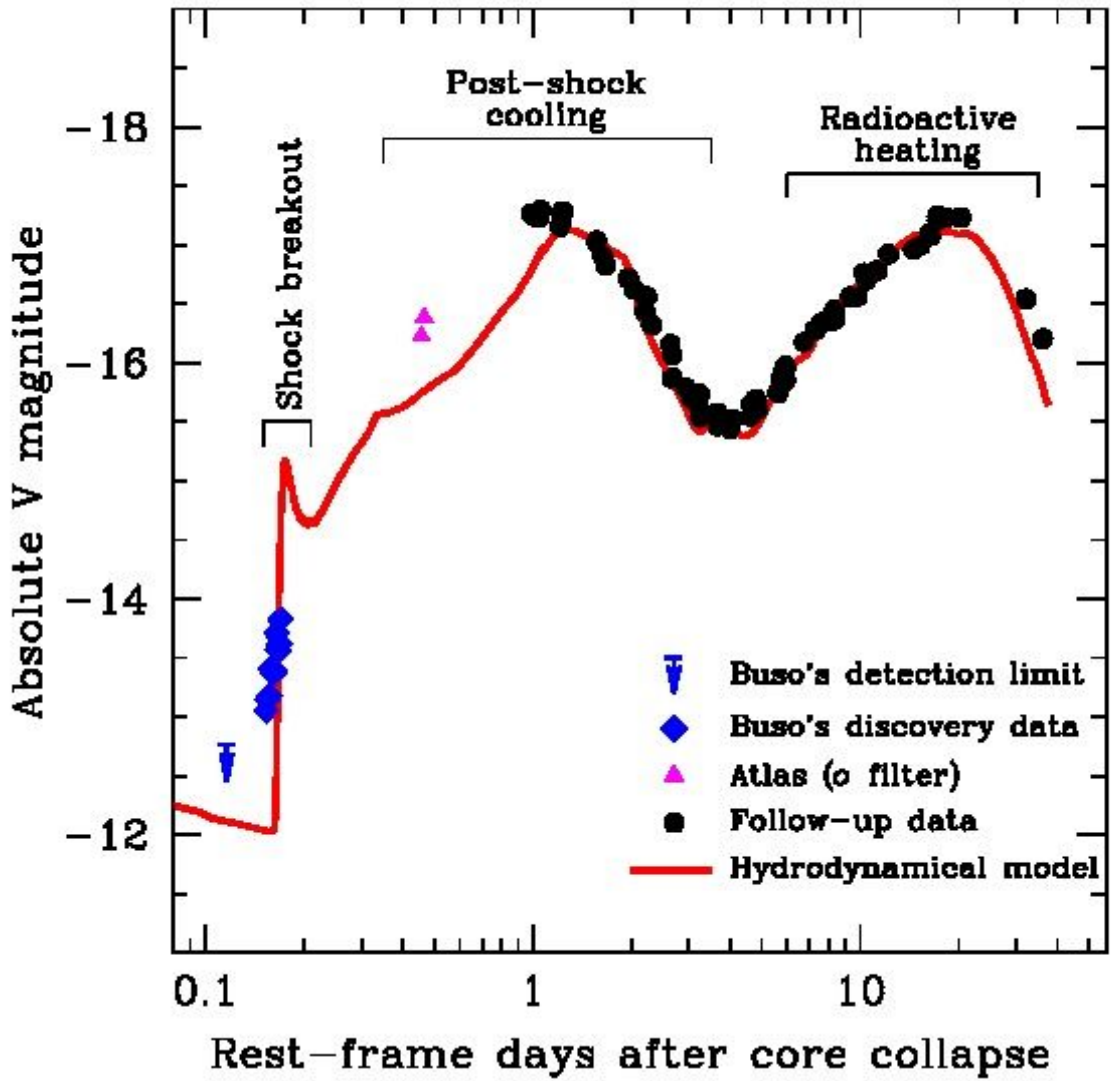

**Hydrodynamical model of the V-band light curve of SN 2016gkg.** Our preferred model (red line) is able to self-consistently reproduce the observations (points) during three distinct phases of the SN evolution, with different characteristic timescales and dictated by different physical properties: the SBO phase, the post-shock cooling peak, and the radioactivity-powered peak. Note the use of a logarithmic scale for the time axis.

**Methods**

**SN discovery data.** SN 2016gkg was discovered on September 20, 2016 [10] by amateur astronomer Víctor Buso from Rosario, Argentina, using a 406 mm Skywatcher Newtonian f=4.4 reflector equipped with a ZWO ASI1600 MM-C camera and a Clear (L) filter. Observations comprised four series of 40, 17, 20, and 21 images, each with an exposure time of 20 s. Images were bias- and dark-subtracted, flat-fielded, and aligned using the MaxIm DL software. The SN is visible in the three last series of images at α = 01h34m14s:46 and δ = −29°26'25".0 (J2000), whereas a stack of the first 40 images shows no evidence of the SN. Clear-band photometry was calibrated to standard V-band magnitudes based on five nearby stars in the AAVSO Photometric All-Sky Survey (APASS) catalogue [31]. The location and catalog magnitudes of the comparison stars are given in Extended Data Table 1 and Extended Data Figure 1. We decided to transform magnitudes to the *V* band given the dense follow-up coverage in that band (including Swift/UVOT observations), as compared with other bands that also lie within the range of the Clear filter. Additionally, as explained further below, the Clear band is centered similarly to the *V* band, which reduces the transformation error to a minimum. The results of the photometry measurements described in this section are listed in Extended Data Table 2. We note that the airmass during the complete observing time ranged between 1.00 and 1.03; thus, we expect no large systematic error in the photometry produced by differences in colour between the SN and the comparison stars.

We estimated the detection limit on the stack of 40 images from the first series. Following [32], we set the 5-sigma detection limit at the magnitude level where point sources are detected with a 50% probability. To find this magnitude, we first calibrated the V -band zeropoint of the combined image using the DAOPHOT photometry of the comparison stars. Then, we added artificial point sources in an area around the SN location, with varying apparent magnitudes in the range of $19.0 < V < 20.0$ mag. We did this in groups of 1000 artificial stars within each 0.1 mag bin. We finally used the "daofind" task in the DAOPHOT package of IRAF [33] to recover the artificial stars and thus determined the recovery fraction at each magnitude bin. We set the detection threshold $Q$ in "daofind" as a factor of 3–4 times the background noise level. This was based on the image scale value of $p = 0.85$ that results from the measured point-source full width at half-maximum intensity (FWHM) of 2.1 pixels in the combined image [see Figure 2 in 32]. Applying threshold values between 3 and 4, we obtained consistent fractions of recovered artificial sources as a function of magnitude. The resulting detection limit was $V \approx 19.4$ mag, which is about 4.5 mag fainter than the brightness at both the cooling and nickel peaks.

In order to increase the signal-to-noise ratio in the last three series of images, we combined them into several groups. On the one hand, we produced a single combined image per series. On the other hand, we computed eleven averaged images from groups of five or six individual exposures (the number varied in each group owing to the different number of images in each series). Extended Data Figure 2 shows a mosaic of the rising SN as seen in the series of combined images. We performed point-spread-function (PSF) photometry of the SN using the DAOPHOT package. Photometric zero points were computed for each individual and combined image, using the five comparison field stars. The results are listed in Extended Data Table 2 and shown in Figure 1 of the main text. The quick initial rise is evident both from the individual-image and combined image photometry. A linear fit to the resulting light curves yielded slopes of $37.3 \pm 5.3$ mag d$^{-1}$ using the photometry from individual exposures, $43.4 \pm 6.1$ mag d$^{-1}$ using eleven data points, and $47.2 \pm 7.8$ mag d$^{-1}$ using three data points. The reduced $\chi^2$ is about unity in all cases, which indicates that no significant departures from a linear rise in magnitudes was seen. The relatively lower slope obtained from individual measurements, although statistically compatible, can be explained by the fact that the SN is initially very near the detection limit of individual images. This introduces a bias toward brighter measurements in DAOPHOT, which in turn causes the slope to be slightly smaller. Combined images into groups of five or six show the SN comfortably above the detection limit. The corresponding slope is thus more reliable in these cases.

We tested our results by performing aperture photometry. We set the aperture size to approximately the value of the source's FWHM. Although the results presented a slight systematic difference during the second series (when the SN is the faintest), in the sense that aperture photometry yielded brighter magnitudes, the main result was confirmed: the SN brightened by about 40 mag d$^{-1}$. Other possible sources of error in the photometry are contamination by background host-galaxy light, and unaccounted variability of the PSF shape across the image. The host galaxy is faint and relatively smooth at the SN location (see Extended Data Figure 1), and thus is expected to be accurately subtracted during the measurement of the SN count rate. Moreover, the image quality is very stable throughout the complete observations (see FWHM values in Extended Data Table 2). Therefore, any contamination would be approximately constant in time, hence not affecting the observed slope. Regarding the PSF variability, we found its shape to be consistent across the image tested its shape across the image. Also, the SN and comparison stars are all located near the centre of the image, where the PSF shape is very stable.

We also tested the possible effect on the measured slope that would be caused by a varying colour of the SN during the observations. At very early times, the temperature evolution can be very fast, thus changing the SN colour. Since we transform the measured counts in the Clear band to the *V* band, a rapid change of colour can affect the derived magnitudes. Beyond a constant error introduced by the fact that the comparison stars are expected to be redder than the SN, the effect on the slope would arise solely from the change in SN colour. The SN colour during this phase is unknown. However, our models (see **Hydrodynamical models** below) show that the SN peaks at a temperature of about $2 \times 10^5$ K soon after explosion, and cools off to about $5 \times 10^4$ K a few hours later. Assuming black-body spectra of varying temperature in the range $(20–200) \times 10^3$ K, we estimated the possible variation in the calibrated *V*-band SN photometry. For this purpose we calculated an approximate Clear-band transmission function, including contributions from the manufacturer's filter transmission, the quantum efficiency of the detector, two aluminum reflecting surfaces, and the atmospheric transmission. This filter passband is available online (see **Data Availability**). Synthetic (Clear – *V*) colours varied by less than 0.05 mag in the range of temperatures considered above. Such a variation is less than 10% of the observed change in magnitude, which is below the uncertainty in the fitted slope. This means that the observed rise cannot be caused by an increasing amount of flux entering the optical range as the SN cools down. We also estimated the size of the error introduced by the fact that the comparison stars are substantially redder than the SN during this phase (see Extended Data Table 1). On the basis of those colours we assumed that the stars are of spectral type between G2 and K3 and adopted ATLAS99 atmosphere models to represent their SEDs. Synthetic colour differences between these stars and the SN were below 0.1 mag for any assumed SN temperature and stellar spectrum. The error is small because the Clear passband, although much wider, is centered near the *V*-band central wavelength. The extra flux introduced by the SN on the blue side of the filter passband is compensated by extra flux from the stars on the red side. The correction from Clear to *V* is thus accurate enough to the purposes of the current analysis.

**Light curves.** V. Buso and J. L. Sánchez obtained new images of SN 2016gkg on September 21, with Clear, *B*, *V*, and *I* bands. Both observers used identical telescopes and cameras. We performed aperture photometry of the SN and comparison stars, and converted the results to the standard system. The Clear-band data were transformed to *V* magnitudes. Extended Data Table 3 provides those magnitudes.

Follow-up *B*, *V*, *R*, and *I* multiband images of SN 2016gkg were obtained with both the Katzman Automatic Imaging Telescope [KAIT; 34] and the 1 m Nickel telescope at Lick Observatory. All images were reduced using a custom pipeline [35]. Aperture photometry was then obtained from a customised Interactive Data Language (IDL) tool using the IDL Astronomy User's Library Note that owing to the small field of view for both KAIT and the Nickel telescope, we were only able to use one appropriate nearby star from the APASS catalogue as the reference star for calibration. Its magnitudes were first transformed into the Landolt system using the empirical prescription presented by Robert Lupton [36], and then transformed to the KAIT and Nickel natural systems. Apparent magnitudes were all measured in the KAIT4/Nickel2 natural system [35, 37]. The final results were transformed to the standard system using local calibrators and colour terms for KAIT4 as given in Table 4 of [35], and updated Nickel colour terms as given by [37]. Extended Data Table 3 lists the resulting standard-system magnitudes from the KAIT and Nickel telescopes.

Extended Data Figure 3 shows the resulting light curves of SN 2016gkg compared with those of the well-observed Type IIb SN 1993J and SN 2011dh. In the figure we also added the early-time *V*-band (or transformed) photometry from ATLAS, ASAS, LCOGT, and Swift.

**Spectra.** Spectroscopic observations of SN 2016gkg were performed using the Kast spectrograph on the Lick Observatory 3 m Shane telescope on 2016 September 24, November 3, December 4, and December 23, the Low Resolution Imaging Spectrometer [LRIS; 38] on the 10 m Keck-I telescope on 2016 September 28 and 2017 January 2, and the Deep Imaging Multi-Object Spectrograph [DEIMOS; 39] on the 10 m Keck-II telescope on 2016 October 25 with both the 600 and 1200 line mm$^{-1}$ gratings. Data were obtained at the parallactic angle [40] to ensure accurate relative spectrophotometry; moreover, LRIS is equipped with an atmospheric dispersion corrector. Standard data reduction (including bias subtraction, flat-fielding, and spectral extraction) was performed within IRAF. The spectra were flux calibrated via observations of spectrophotometric standard stars at similar airmasses to the SN observations. The spectra are shown in Extended Data Figure 3. Also displayed for comparison are spectra of the Type IIb SNe 1993J [41, 42] and 2011dh [43]; these spectra were all obtained from WISeREP [28]. We find, generally, that the SN 2016gkg spectra bear a stronger resemblance to those of SN 2011dh than to those of SN 1993J.

We produced synthetic spectra using the SYNOW code [see 44, and references therein] with the aim of estimating expansion velocities as a function of phase. SYNOW provides approximate continuum and line-strength levels, based on simple assumptions. It can, however, provide a robust estimate of the velocity at the photosphere from the overall fit to the observed spectrum including lines of a number of ionic species. Each SYNOW spectrum was computed consistently with the rest of the epochs, by keeping a smooth variation of the photospheric temperature and velocity with epoch. The derived photospheric velocities, shown in panel b of Extended Data Figure 4, were used to compare with the hydrodynamical models.

**Hydrodynamical model.** To analyse the SN photometry and photospheric velocity evolution, we used a one-dimensional Lagrangian hydrodynamics code that assumes flux-limited radiation diffusion for optical photons, and a one-group approximation for the nonlocal deposition of rays produced by radioactive decay [45]. The code simulates the SN explosion by injecting energy in a hydrostatic structure and self-consistently following the shock-wave propagation inside the star, the shock breakout, and the subsequent expansion of the SN ejecta during the photospheric phase. For the present work the code was adapted to include light-travel time effects and limb-darkening corrections, following the prescription of [4], which are only relevant at the earliest epochs (the first 2 h of the supernova evolution.

As initial configurations (pre-SN models) we employed hydrostatic structures from single stellar evolution calculations. Specifically, models with pre-SN masses of 3.3 $M_\odot$ (HE3.3), 4 $M_\odot$ (HE4), 5 $M_\odot$ (HE5), and 6 $M_\odot$ (HE6), which correspond to initial main-sequence masses of 13, 15, 18, and 20 $M_\odot$ (respectively), were tested [46]. All these configurations are compact hydrogen-free structures with radii of R < 3 $R_\odot$, and were evolved from He burning until core collapse, assuming solar initial abundances [46]. However, these pre-SN models were later modified to take into account the presence of a tenuous hydrogen envelope (of ≤1 $M_\odot$). This model construction is proven to be necessary to successfully reproduce the two-peak morphology of the light curves and the spectra of Type IIb SNe [see more details in 23].

Our first step in the modelling was to find a set of parameters, such as explosion energy (E), ejected mass ($M_{ej}$), and the mass of synthesised $^{56}$Ni ($M_{Ni}$) and its distribution, that provided a good representation of the main bolometric light-curve peak and the photospheric velocity ($v_{ph}$) evolution. Owing to large uncertainties in distance, reddening, bolometric corrections, photospheric velocities, etc., we do not attempt a statistical fit to the observations. However, our conclusions are not affected by this. The model that provides the best overall agreement with the bolometric light curve and expansion velocities, shown with a solid line in Extended Data Figure 4 (panels a and b), corresponds to the HE5 model for an explosion energy of E = 1.2 × 10$^{51}$ erg, a $^{56}$Ni mass of $M_{Ni}$ = 0.085 $M_\odot$, and an ejecta mass of $M_{ej}$ = 3.4 $M_\odot$, assuming the formation of a compact remnant ($M_{cut}$) of 1.6 $M_\odot$. The figure also shows models with lower and higher mass that resulted in worse fits to the data. Note that the HE4 model, which corresponds to Mej = 2.5 $M_\odot$, E = 1 × 10$^{51}$ erg, $M_{Ni}$ = 0.087 $M_\odot$, and $M_{cut}$ = 1.5 $M_\odot$, provides a possible solution, although slightly worse than that of HE5. Indeed, models having intermediate parameters between those of HE4 and HE5 —with $M_{ej}$ = 2.5 – 3.4 $M_\odot$, E = (1–1.2) × 10$^{51}$ erg, and $M_{Ni}$ = 0.085 – 0.087 $M_\odot$— are also valid. Therefore, these values can be considered as the ranges of validity for the physical parameters. An important conclusion of this analysis is that the progenitor of SN 2016gkg needs to be a relatively low-mass He star, as is commonly the case for stripped-envelope SNe [47, 48, 49].

Once the global parameters were set, we focused on the modelling at epochs prior to the onset of radioactive-decay domination. At such times, the light curve is extremely sensitive to the extent (radius) and mass of the H-rich envelope. Therefore, we modified the HE5 model initial structure by smoothly attaching a low-mass H-rich envelope in hydrostatic and thermal equilibrium. Note that the HE4 model could also have been used in this analysis without changing the conclusions. Different configurations were tested with various progenitor radii and envelope masses. By comparing with the observations, we were able to find our preferred configuration, denoted as the preferred model and shown in Figure 3 in the main text, corresponding to an object with an H-rich envelope of radius R = 320 $R_\odot$ and a mass of $M_{env}$ = 0.01 $M_\odot$. The preferred model provides a satisfactory match to the observations. Although the agreement during the post-shock cooling peak is not perfect, it should be noted that the model consistently reproduces the complete SN evolution. Panel c of Extended Data Figure 4 shows this model with a solid line, compared with models of larger (smaller) progenitor radii, which overestimate (underestimate) the luminosity during the post-shock cooling phase. From these comparisons we conclude that reasonable ranges of validity for the radius and mass of the envelope are R = 300 – 340 $R_\odot$, and $M_{env}$ = 0.01 – 0.09 $M_\odot$. Note that this analysis was based on the V-band light curve instead of the bolometric luminosity. The reason for using the V-band data was that the earliest observations were done with that band or with a Clear filter that was transformed to the V band. In addition, a bolometric correction at such early epochs is highly uncertain. Therefore, we computed theoretical V-band photometry assuming a black-body spectral energy distribution at the thermalisation temperature [23].

Previous analyses of the post-shock cooling emission of SN 2016gkg arrived at different values of R and $M_{env}$, for example, $R \approx 257 R_\odot$ (ref. 16), $R = (50–125) R_\odot$ (ref. 17) and $R = (44–131) R_\odot$ (ref. 18). In these analyses an analytical approach was used, which is simpler than the hydrodynamical modelling presented here. Moreover, such analytic approaches have been proven to be wrong in the interpretation of the progenitor radius for the similar type IIb supernova SN 2011dh[23,46]. On the contrary, a similar hydrodynamical model for the post-shock cooling emission[47] of SN 2016gkg yielded a radius of $R \approx 200 R_\odot$ and an envelope mass of $M_{env} \approx 0.02 M_\odot$, which are smaller than our values. Although the hydrodynamical code used[47] is similar to ours, the initial structures are not: instead, parametric models were used that allow the modification of the initial density profiles to take into account the existence of a tenuous extended envelope. This method, contrary to ours, does not ensure hydrostatic and thermal equilibrium of the external envelope. This could be a cause of the differences in the derived parameters. Several interesting conclusions can be derived from Extended Data Fig. 4, as follows.

(1) Our model indicates that Type IIb SN optical light curves show three peaks, rather than the double peaks usually referred to in the literature. These are the SBO peak, the post-shock cooling peak (CP), and the nickel-powered peak (see also Figure 3 in the main text).
(2) All models indicate that the discovery data can be interpreted as the SBO, regardless of the adopted progenitor radius. This is still valid for the values proposed by other authors [18, 51].
(3) The progenitor radius has a much more noticeable effect on the decline rate than on the rise rate during the CP.
(4) The larger the progenitor radius, the more luminous and later the CP becomes.

To further test our SBO interpretation of the early-time data, we extensively explored parameters other than the radius that could increase the slope of the CP to a similar rate as that of the SBO peak (i.e., ~ 40 mag d$^{-1}$). We found that the explosion energy has the strongest effect on increasing this slope (see panel d of Extended Data Figure 4). However, within the range of E values allowed by the modelling of the complete SN evolution, the CP rise is always appreciably slower than the SBO rise. This is true even in an extreme case, with an explosion energy of $5 \times 10^{51}$ erg (~4 times larger than our preferred model), as shown in Extended Data Figure 5. Our analysis demonstrates that the initial rise is always steeper than the rise to the post-shock cooling peak, and that there is always a local maximum in the light curve between both phases, provided realistic pre-SN structures are assumed. This result gives support to our SBO interpretation.

Remarkably, we proved the existence of a model that is able to reproduce three distinct light-curve phases with very dissimilar timescales (note the logarithmic timescale in Figure 3 of the main text), assuming a standard set of parameters that fit normal SNe IIb. Nevertheless, a close look at the earliest phases shows that the observed rise is slower than that of the models during the SBO phase (see Extended Data Figure 5). We thus tested if this situation could be ameliorated by adding some extra surrounding material to the initial density structure. We did not assume this material to be in hydrostatic equilibrium. It could be material ejected by the progenitor prior to the SN explosion, probably in the form of a dense wind. Extended Data Figure 5 shows a model with such circumstellar material (CSM; dashed line). Clearly, the presence of this material slows the rise during the SBO, in better concordance with the observations, without affecting the light curve at later epochs (at times ≥ 1 day) owing to the small amount of matter involved. Specifically, the model presented here corresponds to a mass of 0.002 $M_\odot$ distributed out to $3 \times 10^{13}$ cm, assuming a steep power-law density profile with an index of 10. This corresponds to an average mass-loss rate of $6 \times 10^{-4}$ $M_\odot$ yr$^{-1}$, for a wind velocity of $v_{wind}$ = 100 km s$^{-1}$. However, we found that the exact properties of this material are not very relevant. In fact, assuming a range of other density structures with different slopes and extensions produces similar results. In particular, for a constant wind profile (with an index of r$^{-2}$) we found almost the same results, but in this case the mass-loss rate increased by almost two orders of magnitude. Based on our tests, we can say that the mere presence of this material is enough to slow down the SBO rise, with very little dependence on its exact nature. In this context, it is noteworthy that recently there is increasing evidence from earlytime observations of normal SNe (photometry and spectroscopy) of the existence of surrounding material in the vicinity of the progenitor, possibly produced by a dense wind or an eruption that occurred shortly before the explosion [13, 15, 52, 53].

It has been noticed that nonthermal processes could play a role to shape the light curve during the SBO phase [25]. This second-order effect could smooth the SBO peak. However, according to those calculations, the initial rise rate remains basically unchanged. This suggests that, even if deriving detailed CSM properties from our models would be premature, the conclusion of the SBO signal detection would remain well-founded. A deeper analysis of these effects is thus left for future study.

**Previous SBO claims.** Some SNe have been associated in the past with possible SBO emission. The outstanding cases of SN 2006aj and SN 2008D [11, 12, 54, 55] are worth noting. SN 2006aj was connected with a gamma-ray burst (GRB), and SN 2008D was preceded by an X-ray flash. The X-ray transient in both cases was interpreted by some authors as being produced by the SBO. However, the peculiar characteristics of both events and the lack of a model that fully described the hard and soft emission cast some doubt on this interpretation [55, 56, 57]. Their early-time optical data are shown in Figure 2 of the main text. The initial optical rise of SN 2008D has a slope and luminosity similar to those of SN 2016gkg during the cooling peak. Interestingly, Swift/UVOT *V*-band data of SN 2006aj obtained within 1 hr after the associated GRB and presented by [11] show a similarly steep rise of 49 ± 16 mag d$^{-1}$ as that of SN 2016gkg at discovery, although with a much larger luminosity. Reanalysis of the early-time data based on the 2015 measurements available in the Swift Optical/Ultraviolet Supernova Archive [SOUSA; 58, and P. Brown, private communication] provides a smaller slope of 21 ± 12 mag d$^{-1}$. Nevertheless, the behaviour of this emission may still be interpreted as an SBO, although emerging from some CSM instead of the stellar surface. It should also be noted that the rise slope of the SBO in this case may be reduced by a declining contribution from the GRB afterglow.

More recently, the case of KSN 2011d discovered by the Kepler mission was pointed out as an SBO detection based on an excess in the early-time optical light curve relative to a simple analytic model [14]. However, as indicated by [59], modifying the data binning and comparison function shows that there is no statistical significance for an SBO in KSN 2011d.

**Progenitor candidate.** The SN 2016gkg site is contained in publicly available archival HST images, obtained with the Wide-Field Planetary Camera 2 (WFPC2) in bands F450W, F606W, and F814W on 2001 August 21, as part of program GO-9042 (PI S. Smartt). We also obtained images of the SN itself on 2016 October 10 with HST, with theWide-Field Camera 3 (WFC3) UVIS channel in band F555W, as part of the Target of Opportunity (ToO) program GO-14115 (PI S. Van Dyk). The observations consisted of 24 dithered frames, each of 10 s duration; the short exposure time mitigated against possible saturation by a potentially bright SN (we knew that the SN brightness was increasing at the time, but not to what level). The frames were combined into a final mosaic of 240 s total exposure using AstroDrizzle [60] in DrizzlePac [61] within PyRAF. Adopting 13 stars in common between the WFPC2 image mosaic at F606W and the WFC3/UVIS mosaic at F555W, we astrometrically registered the two datasets with a root-mean-square (RMS) uncertainty of 0.42 WFPC2/WF pixel (0".042, see the registered images in Extended Data Figure 6). After measuring the centroid of the SN in the WFC3 mosaic, we found that the SN position on the WFPC2 mosaic is then (1552.47, 196.39). On the WFPC2 mosaic we measured a centroid for the progenitor candidate of (1552.63, 196.06). This is a difference of 0.37 pixel, which is within the astrometric uncertainty. So, we consider this progenitor candidate to be solidly identified; it will not be until the candidate has vanished well after the SN explosion that its identity will have been confirmed with little doubt. We note that both [16] and [18] possessed more ambiguity in their identification of the candidate.

We measured photometry for the progenitor candidate running Dolphot [62] on the individual "c0m" WFPC2 frames after masking cosmic rays with AstroDrizzle. Although the star is relatively isolated, with little apparent background, we set FitSky=3 (rather than FitSky=1), since the star is only 10 pixels from the edge of the WF4 chip and much of the sky annulus would sit off the edge. We also therefore set RAper=8, as well as InterpPSFlib=1 using the TinyTim PSF library. We enabled charge transfer efficiency corrections by setting WFPC2UseCTE=1 in Dolphot. This resulted in brightnesses on the Vega system of $m_{F450W}$ = 24.07 ± 0.16 mag, $m_{F606W}$ = 24.04 ± 0.07 mag, and $m_{F814W}$ = 23:58 ± 0.14 mag. We note that [16] found 23.60 ± 0.14 mag, 23.72 ± 0.08 mag, and 23.25 ± 0.14 mag, respectively, for their Object A (which we have shown here is the progenitor candidate); these authors also converted the results from [18] from STMAG to VEGAMAG: 23.42, 23.10, and 23.32 mag, respectively. In [16], the authors also noted that this object in the Hubble Source Catalog has 23.85 ± 0.08 mag at F450W and 23.34 ± 0.05 mag at F606W, all VEGAMAG. Our measurements are brighter by 0.3–0.4 mag compared to those of [16]. We have contacted the authors of [16], and we now recognise the source of the differences as being the cosmic-ray masking procedure (they used LACosmic while we used AstroDrizzle), and the combination of the FitSky/RAper parameters in Dolphot (we used 3/8 and they used 1/4).

From our photometry of the progenitor candidate, and assuming a distance modulus of 32.11 ± 0.38 mag and an extinction (Milky-Way only) of $A_V$ = 0.053 mag (see SN site extinction and metallicity in Methods), we obtained absolute magnitudes of $M_{F450W}$ = −8.1 ± 0.4 mag, $M_{F606W}$ = −8.1 ± 0.4 mag, and $M_{F814W}$ = −8.6 ± 0.4 mag. We performed $\chi^2$ fits of the resulting spectral energy distribution (SED) to stellar atmosphere models from ATLAS9 [63] and found best-fit values of $T_{eff}$ = $7250^{+900}_{-850}$ K and luminosity $\log(L/L_\odot)$ = $5.10^{+0.17}_{-0.19}$. The fitted SED is shown in Extended Data Figure 6. Assuming a black body, this corresponds to a progenitor radius of $226^{+98}_{-73}$ $R_\odot$. Such a radius is slightly smaller than (but still compatible with) what we estimated from the hydrodynamical modelling [see 16, 18, 51, for previous estimates].

**Progenitor model.** We attempted to find a consistent progenitor picture based on the information from the light-curve modelling and the pre-explosion photometry. The location of the preexplosion object in the Hertzsprung-Russell diagram (HRD) is not compatible with the endpoints of single stellar evolutionary tracks [see 16], unless some nonstandard enhanced mass loss is assumed. The relatively low progenitor mass derived from the light-curve modelling, however, goes against the possibility of large mass loss produced by an isolated star wind. A more natural scenario is that of a close binary system where the primary star explodes as an SN after transferring mass to its companion. This type of system allows for strong mass loss even in the case of relatively low-mass progenitors. Here we present a possible such scenario for the case of SN 2016gkg. Our proposed model is not supposed to be a unique solution.

We used code described[60] and applied[61] previously to SN 2011dh. This code has detailed and updated physical ingredients (see ref. 61 and references therein). When stars are detached, it works as a standard Henyey code. When the donor star undergoes Roche-lobe overflow[62], the code computes the mass-transfer rate simultaneously with the structure of the donor star in an implicit Henyey-like, numerically stable algorithm. We neglected rotation of the components and assumed that the orbit is circularized and synchronized. We assumed that the accreting star retains a fraction β of the material transferred by the donor component—a free parameter the value of which is kept constant throughout the entire evolution. Here we considered values of $\beta$ = 0.0, 0.25 and 0.50 (that is, non-conservative evolution). The material lost from the system is assumed to have the specific angular momentum of the companion star.

We found good agreement with the observations by assuming a binary progenitor with solar abundance, initial masses of 19.5 $M_\odot$ and 13.5 $M_\odot$, and an orbital period of 70 days. The primary star explodes as an SN with final mass and radius of M = 4.61 $M_\odot$ and R = 183 $R_\odot$, and the final orbital period is 631 days. The total amount of hydrogen remaining in the primary is 6 × $10^{-3}$ $M_\odot$, contained in the outer ~ 0.06 $M_\odot$ of the star. The surface abundance is $X_{surf}$ = 0.21. The model stays inside the error box of Extended Data Figure 6 (panel d) for the final 14,000 yr of evolution. All of these values are almost independent of the uncertain value of β.

Qualitatively, the evolution of the progenitor of SN 2016gkg is very similar to that of SN 2011dh. For both objects, binary models are more plausible candidates than single stars, because isolated objects need very specific mass-loss rates to account for the final luminosity and effective temperature (L and $T_{eff}$) indicated by the pre-supernova observations. In binary systems, the donor (progenitor) star spends almost all of its final nuclear burning stages (carbon, neon, oxygen and silicon), which last for several thousand years, under Roche-lobe-overflow conditions. This places the progenitor in a well-defined region of the HRD (Extended Data Fig. 6), inside the error box in ($L$, $T_{\text{eff}}$) for SN 2016gkg. Thus, binary systems provide a natural reference frame for interpreting the evolution of the progenitor of SN 2016gkg.

We note that [18] proposed a very different progenitor with 15 $M_\odot$ + 1.5 $M_\odot$ and $P_{\text{ini}}$ = 1000 days, which yielded a presupernova mass of 5.2 $M_\odot$. Their evolutionary track is completely different from ours, since it undergoes core He burning as a red supergiant, and after He exhaustion it executes very large loops. Their scenario requires some fine tuning of the initial conditions for the presupernova model to be at the observed location in the HRD; as discussed above, ours does not.

Remarkably, for the computed systems, most of the mass accreted by the companion star is gained before core He burning. Thus, there is plenty of time for the accreted mass to accommodate to the stellar structure. This implies that the companion star remains close to the zero-age main sequence in the HRD, while being somewhat overluminous for its mass. Similar to the case of SN 2011dh, our calculations predict the existence of a hot companion to the progenitor that should remain after the explosion. The position of the companion star in the HRD is dependent on its final mass, and thus on β. At the moment of the explosion, the companion star is still undergoing core hydrogen burning. The presence of this object may be tested with future observations, once the SN fades sufficiently

**SN site extinction and metallicity.** In order to estimate a colour excess for SN 2016gkg, we compared its (B – V) colours with those of SN 2011dh, which shows very similar spectral evolution (see Extended Data Figure 3). Adopting only the Galactic reddening of E(B – V) = 0.017 mag for SN 2016gkg, and a total reddening of E(B – V) = 0.074 mag for SN 2011dh [43], the colour curves of both SNe match very well. This indicates that the host-galaxy reddening for the former SN is negligible.

In order to test this, we inspected the optical spectra for signatures of dust extinction by looking for interstellar absorption lines. In our highest-resolution spectrum obtained with DEIMOS on 2016 October 25, we detected the Na I D doublet both from the Milky Way and at the redshift of the host galaxy. The equivalent width (EW) of the D1+D2 lines was 0.16 + 0.13 Å for the Galactic component and 0.43 + 0.26 Å for the host-galaxy component. This may indicate a larger host extinction than that from the Milky Way [as assumed in 17]. However, the Na I D EW has been shown to be a poor indicator of dust extinction [67]. Following [67], we studied instead the diffuse interstellar band (DIB) at 5780 Å. Such a feature is not detected in our DEIMOS spectrum, with a limiting EW of ~0.01 Å. This is indicative of a low host-galaxy extinction, $A_V$ < 0.05 mag.

Based on the colour comparison with SN 2011dh and on the study of spectral lines, we decided to neglect host-galaxy extinction. In any case, our results are not affected by this assumption.

We can also estimate the metallicity more directly from the spectrum of an H II region, at α = 01h34m14.53s, δ = −29°26'16".4 (J2000), which is ~8".6 nearly due north of the SN site [68]. We had included the H II region in the slit, while observing the SN with Keck/DEIMOS on 2016 October 25 with the lower-resolution grating. We note that, owing to the short DEIMOS slit length, for strong emission lines (in this case, H) not much spatial area exists on the spectral image for accurately estimating the overall night-sky value, likely introducing systematic uncertainty in the strong-line flux. Nonetheless, we measured the

Balmer decrement from the observed spectrum and estimate that $A_V$ = 3.5 mag for the nebula. The line-of-sight Galactic foreground contribution to the extinction is comparatively low, $A_V$ = 0.053 mag [69]. A high extinction is plausible, given the conspicuous presence of a counterpart of the H II region in archival Spitzer Space Telescope data. The nebula corresponds to a luminous source at both 3.6 and 4.5 µm, and one of the brightest sources at 24 µm, in the outer disk of the galaxy. We corrected the spectrum for this extinction, as well as for an assumed recession velocity of 1481 km s$^{-1}$ (from the NASA/IPAC Extragalactic Database, NED), and show the corrected spectrum in Extended Data Figure 6.

We enlisted the various strong-line indicators used to estimate the metallicity of extragalactic H II regions; these lines are labelled in panel e of the Extended Data Figure 6. We measured their fluxes from the corrected spectrum and list them in Extended Data Table 4. Unfortunately, the spectrum did not go blueward enough where we could employ the well-calibrated $R_{23}$ indicator or [N II]/[O II], which depends on the intensity of the [O II] λ3727 line [e.g., 70]. Instead, we had to rely on other indicators. From the indicators $R_3$, $N_2$, and $S_2$, as defined by [71] and using the "S calibration" (their Equation 6), we find that 12 + log (O/H) = 8.65. Considering together the indicators $R_3$, $N_2$, and $O_3N_2$ from [72], derived with their online tool at http://www.arcetri.astro.it/metallicity/, we arrive at 12+log (O/H) = 8.7.

Culling all of these estimates and assuming a solar abundance 12 + log (O/H) = 8.69 ± 0.05 [73], the metallicity at the SN 2016gkg site appears to be consistent with solar, and we have adopted this throughout.

**Data availability.** The datasets analysed during the current study are available from the following website: http://fcaglp.unlp.edu.ar/~gaston/data/sn2016gkg/.
**Code availability.** We have opted not to make the SN light-curve modelling code nor the binary evolution code available because they have not been prepared for portability, and they lack the necessary documentation for general use. However, all optical spectra will be made available at WISeREP [28].

**Extended Data Table 1:** Comparison Stars in the Field of SN 2016gkg

| Star | α (J2000.0) [hh:mm:ss] | δ (J2000.0) [°:′:″] | B | V | R | I |
|---|---|---|---|---|---|---|
| 138 | 01:34:29.09 | -29:29:40.3 | 14.741 ± 0.038 | 13.828 ± 0.012 | 13.402 ± 0.033 | 13.003 ± 0.045 |
| 143 | 01:34:07.46 | -29:21:37.8 | 15.015 ± 0.013 | 14.293 ± 0.006 | 13.886 ± 0.103 | 13.504 ± 0.146 |
| 151 | 01:33:55.52 | -29:20:18.1 | 16.091 ± 0.081 | 15.110 ± 0.045 | 14.722 ± 0.080 | 14.357 ± 0.104 |
| 155 | 01:34:08.51 | -29:29:04.4 | 16.376 ± 0.110 | 15.464 ± 0.030 | 15.001 ± 0.034 | 14.568 ± 0.038 |
| 161 | 01:34:02.36 | -29:24:03.3 | 16.694 ± 0.056 | 16.050 ± 0.017 | ... | ... |

**Extended Data Table 2:** Discovery Imaging Description and Photometry of SN 2016gkg

| Image number | JD - 2457651.0 [day] | Exp. time [s] | V [mag] | FWHM ["] | zero point [mag] | Image number | JD - 2457651.0 [day] | V [mag] | FWHM ["] | zero point [mag] |
|---|---|---|---|---|---|---|---|---|---|---|
| 1 - 40 | 0.69693184 | 800 | > 19.4 | 1.93 ± 0.13 | 1.023 ± 0.011 | 74 | 0.74384259 | 19.01 ± 0.20 | 2.20 ± 0.11 | 2.925 ± 0.008 |
| 41 - 57 | 0.73615480 | 340 | 19.03 ± 0.08 | 1.61 ± 0.09 | 1.235 ± 0.018 | 75 | 0.74414352 | 19.04 ± 0.58 | 2.18 ± 0.18 | 2.954 ± 0.013 |
| 41 - 46 | 0.73452949 | 120 | 19.11 ± 0.13 | 1.59 ± 0.10 | 1.219 ± 0.013 | 76 | 0.74446759 | 18.63 ± 0.30 | 2.82 ± 0.44 | 3.071 ± 0.020 |
| 47 - 52 | 0.73629636 | 120 | 19.02 ± 0.13 | 1.67 ± 0.15 | 1.220 ± 0.018 | 77 | 0.74476852 | 18.58 ± 0.20 | 2.10 ± 0.15 | 2.749 ± 0.010 |
| 53 - 57 | 0.73793518 | 100 | 18.98 ± 0.12 | 1.57 ± 0.12 | 1.222 ± 0.020 | 78 - 98 | 0.74817347 | 18.48 ± 0.05 | 2.06 ± 0.07 | 2.780 ± 0.016 |
| 41 | 0.73376157 | 20 | 18.95 ± 0.39 | 1.71 ± 0.20 | 1.244 ± 0.023 | 78 - 83 | 0.74595499 | 18.46 ± 0.08 | 2.28 ± 0.08 | 2.845 ± 0.012 |
| 42 | 0.73410880 | 20 | 19.55 ± 0.36 | 2.21 ± 0.24 | 1.332 ± 0.021 | 84 - 88 | 0.74763662 | 18.60 ± 0.09 | 2.00 ± 0.08 | 2.814 ± 0.025 |
| 43 | 0.73439815 | 20 | 18.95 ± 0.19 | 1.53 ± 0.35 | 1.248 ± 0.015 | 89 - 93 | 0.74904883 | 18.55 ± 0.09 | 2.00 ± 0.06 | 2.780 ± 0.014 |
| 44 | 0.73468750 | 20 | 18.77 ± 0.26 | 1.53 ± 0.21 | 1.220 ± 0.010 | 94 - 98 | 0.75049746 | 18.34 ± 0.08 | 2.00 ± 0.07 | 2.783 ± 0.018 |
| 45 | 0.73496528 | 20 | 19.65 ± 0.38 | 1.67 ± 0.38 | 1.199 ± 0.013 | 78 | 0.74512731 | 18.70 ± 0.19 | 2.06 ± 0.10 | 2.708 ± 0.012 |
| 46 | 0.73525463 | 20 | 19.41 ± 0.42 | 1.56 ± 0.28 | 1.238 ± 0.016 | 79 | 0.74550926 | 18.39 ± 0.14 | 2.48 ± 0.28 | 2.866 ± 0.016 |
| 47 | 0.73554398 | 20 | 19.01 ± 0.31 | 2.13 ± 0.46 | 1.165 ± 0.009 | 80 | 0.74582176 | 18.80 ± 0.31 | 2.35 ± 0.14 | 2.953 ± 0.011 |
| 48 | 0.73585648 | 20 | 19.26 ± 0.41 | 1.53 ± 0.52 | 1.221 ± 0.020 | 81 | 0.74612269 | 18.76 ± 0.35 | 2.89 ± 0.44 | 3.103 ± 0.008 |
| 49 | 0.73614583 | 20 | 19.14 ± 0.33 | 1.71 ± 0.29 | 1.201 ± 0.015 | 82 | 0.74642361 | 18.11 ± 0.15 | 2.22 ± 0.10 | 2.770 ± 0.009 |
| 50 | 0.73644676 | 20 | 18.45 ± 0.12 | 2.07 ± 0.27 | 1.274 ± 0.022 | 83 | 0.74672454 | 18.47 ± 0.17 | 2.22 ± 0.15 | 2.741 ± 0.026 |
| 51 | 0.73673611 | 20 | 19.40 ± 0.39 | 1.84 ± 0.43 | 1.208 ± 0.023 | 84 | 0.74706019 | 18.43 ± 0.18 | 2.12 ± 0.12 | 2.770 ± 0.026 |
| 52 | 0.73704861 | 20 | 19.59 ± 0.50 | 1.64 ± 0.55 | 1.174 ± 0.014 | 85 | 0.74737269 | 18.67 ± 0.16 | 1.97 ± 0.09 | 2.756 ± 0.014 |
| 53 | 0.73734954 | 20 | 19.17 ± 0.24 | 1.52 ± 0.27 | 1.166 ± 0.015 | 86 | 0.74765046 | 18.84 ± 0.31 | 1.97 ± 0.15 | 2.769 ± 0.017 |
| 54 | 0.73765046 | 20 | 19.23 ± 0.30 | 1.77 ± 0.23 | 1.255 ± 0.014 | 87 | 0.74791667 | 18.55 ± 0.15 | 2.81 ± 0.30 | 2.979 ± 0.025 |
| 55 | 0.73793982 | 20 | 18.67 ± 0.20 | 1.63 ± 0.51 | 1.197 ± 0.013 | 88 | 0.74818287 | 18.43 ± 0.15 | 1.78 ± 0.16 | 2.753 ± 0.017 |
| 56 | 0.73822917 | 20 | 19.02 ± 0.29 | 1.58 ± 0.42 | 1.218 ± 0.024 | 89 | 0.74843750 | 18.63 ± 0.16 | 2.01 ± 0.09 | 2.777 ± 0.014 |
| 57 | 0.73850694 | 20 | 18.90 ± 0.25 | 2.05 ± 0.30 | 1.267 ± 0.019 | 90 | 0.74870370 | 18.35 ± 0.14 | 1.90 ± 0.10 | 2.756 ± 0.018 |
| 58 - 77 | 0.74201745 | 400 | 18.82 ± 0.07 | 2.03 ± 0.10 | 2.787 ± 0.011 | 91 | 0.74906250 | 18.62 ± 0.17 | 2.38 ± 1.28 | 2.659 ± 0.010 |
| 58 - 62 | 0.73999995 | 100 | 18.75 ± 0.12 | 1.85 ± 0.12 | 2.822 ± 0.014 | 92 | 0.74938657 | 18.19 ± 0.13 | 2.17 ± 0.11 | 2.817 ± 0.012 |
| 63 - 67 | 0.74130100 | 100 | 18.75 ± 0.08 | 2.12 ± 0.09 | 2.825 ± 0.013 | 93 | 0.74965278 | 19.17 ± 0.31 | 2.14 ± 1.17 | 2.684 ± 0.022 |
| 68 - 72 | 0.74262035 | 100 | 18.79 ± 0.10 | 2.18 ± 0.11 | 2.800 ± 0.011 | 94 | 0.74994213 | 18.31 ± 0.14 | 2.10 ± 0.11 | 2.726 ± 0.014 |
| 73 - 77 | 0.74414861 | 100 | 18.79 ± 0.15 | 2.09 ± 0.12 | 2.826 ± 0.012 | 95 | 0.75023148 | 18.43 ± 0.18 | 2.61 ± 0.15 | 2.876 ± 0.023 |
| 58 | 0.73947917 | 20 | 18.59 ± 0.26 | 1.66 ± 0.14 | 2.762 ± 0.016 | 96 | 0.75050926 | 18.23 ± 0.14 | 1.92 ± 0.15 | 2.707 ± 0.021 |
| 59 | 0.73974537 | 20 | 18.93 ± 0.21 | 2.07 ± 0.11 | 2.785 ± 0.016 | 97 | 0.75077546 | 18.51 ± 0.15 | 2.03 ± 0.14 | 2.822 ± 0.020 |
| 60 | 0.74000000 | 20 | 18.82 ± 0.29 | 1.91 ± 0.12 | 2.827 ± 0.016 | 98 | 0.75103009 | 18.32 ± 0.14 | 1.83 ± 0.17 | 2.722 ± 0.023 |
| 61 | 0.74025463 | 20 | 19.20 ± 0.28 | 1.95 ± 0.16 | 2.803 ± 0.014 | | | | | |
| 62 | 0.74052083 | 20 | 18.53 ± 0.20 | 1.79 ± 0.15 | 2.786 ± 0.013 | | | | | |
| 63 | 0.74077546 | 20 | 19.05 ± 0.29 | 2.54 ± 0.11 | 2.926 ± 0.009 | | | | | |
| 64 | 0.74104167 | 20 | 18.35 ± 0.16 | 2.43 ± 0.15 | 2.929 ± 0.028 | | | | | |
| 65 | 0.74129630 | 20 | 19.01 ± 0.23 | 2.01 ± 0.10 | 2.777 ± 0.019 | | | | | |
| 66 | 0.74156250 | 20 | 18.52 ± 0.15 | 2.05 ± 0.07 | 2.772 ± 0.013 | | | | | |
| 67 | 0.74182870 | 20 | 19.04 ± 0.29 | 1.93 ± 0.12 | 2.789 ± 0.011 | | | | | |
| 68 | 0.74208333 | 20 | 18.46 ± 0.16 | 2.32 ± 0.08 | 2.870 ± 0.022 | | | | | |
| 69 | 0.74234954 | 20 | 19.88 ± 0.49 | 2.38 ± 0.27 | 2.888 ± 0.012 | | | | | |
| 70 | 0.74261574 | 20 | 19.32 ± 0.35 | 2.08 ± 0.11 | 2.807 ± 0.012 | | | | | |
| 71 | 0.74287037 | 20 | 18.71 ± 0.21 | 1.85 ± 0.13 | 2.746 ± 0.020 | | | | | |
| 72 | 0.74318287 | 20 | 18.83 ± 0.22 | 2.41 ± 0.07 | 2.825 ± 0.019 | | | | | |
| 73 | 0.74351852 | 20 | 18.94 ± 0.27 | 1.82 ± 0.17 | 2.632 ± 0.018 | | | | | |

**Extended Data Table 3:** Follow-up *BVRI* and Clear Photometry of SN 2016gkg

| MJD | B | V | R | I | Clear | Source |
|---|---|---|---|---|---|---|
| 57652.1 | 15.37 ± 0.35 | 14.94 ± 0.05 | ... | 14.93 ± 0.06 | ... | Buso |
| 57652.3 | ... | 14.95 ± 0.03 | ... | ... | ... | Sánchez |
| 57653.3 | 15.73 ± 0.14 | 15.60 ± 0.08 | 15.57 ± 0.06 | 15.32 ± 0.10 | ... | KAIT |
| 57654.3 | 16.82 ± 0.16 | 16.43 ± 0.12 | 16.23 ± 0.08 | 16.09 ± 0.12 | 16.22 ± 0.03 | KAIT |
| 57658.4 | 16.47 ± 0.08 | 15.89 ± 0.04 | 15.70 ± 0.04 | 15.70 ± 0.06 | 15.74 ± 0.04 | KAIT |
| 57659.4 | 16.34 ± 0.08 | 15.80 ± 0.04 | 15.58 ± 0.04 | 15.56 ± 0.06 | 15.62 ± 0.04 | KAIT |
| 57660.4 | 16.21 ± 0.08 | 15.60 ± 0.04 | 15.43 ± 0.02 | 15.42 ± 0.04 | 15.45 ± 0.05 | KAIT |
| 57661.4 | 15.86 ± 0.04 | 15.33 ± 0.02 | 15.22 ± 0.02 | 15.26 ± 0.04 | ... | Nickel |
| 57662.4 | 15.92 ± 0.12 | 15.38 ± 0.06 | 15.22 ± 0.04 | 15.19 ± 0.06 | 15.22 ± 0.03 | KAIT |
| 57663.3 | 15.84 ± 0.10 | 15.24 ± 0.06 | 15.14 ± 0.04 | 15.02 ± 0.06 | 15.12 ± 0.05 | KAIT |
| 57666.4 | 15.63 ± 0.08 | 15.17 ± 0.04 | 15.01 ± 0.02 | 14.94 ± 0.04 | 14.93 ± 0.02 | KAIT |
| 57667.4 | 15.59 ± 0.06 | 15.07 ± 0.04 | 14.98 ± 0.02 | 14.90 ± 0.04 | 14.93 ± 0.02 | KAIT |
| 57668.4 | 15.39 ± 0.02 | 14.91 ± 0.02 | 14.80 ± 0.02 | 14.74 ± 0.02 | ... | Nickel |
| 57668.4 | 15.53 ± 0.16 | 14.96 ± 0.12 | 14.92 ± 0.08 | 14.81 ± 0.10 | 14.82 ± 0.06 | KAIT |
| 57669.4 | 15.46 ± 0.10 | 14.93 ± 0.06 | 14.85 ± 0.04 | 14.72 ± 0.08 | 14.72 ± 0.04 | KAIT |
| 57671.4 | 15.52 ± 0.10 | 14.93 ± 0.10 | 14.77 ± 0.06 | 14.61 ± 0.10 | 14.70 ± 0.05 | KAIT |
| 57672.3 | ... | ... | ... | ... | 14.71 ± 0.04 | KAIT |
| 57683.3 | 16.72 ± 0.02 | 15.55 ± 0.02 | 15.15 ± 0.02 | 14.92 ± 0.02 | ... | Nickel |
| 57687.3 | 17.10 ± 0.04 | 15.91 ± 0.02 | 15.35 ± 0.02 | 15.05 ± 0.02 | ... | Nickel |
| 57694.3 | 17.46 ± 0.16 | 16.24 ± 0.06 | 15.71 ± 0.04 | 15.29 ± 0.04 | 15.70 ± 0.02 | KAIT |
| 57697.3 | 17.44 ± 0.04 | 16.20 ± 0.02 | 15.68 ± 0.02 | 15.36 ± 0.02 | ... | Nickel |
| 57697.4 | 17.58 ± 0.18 | 16.26 ± 0.08 | 15.73 ± 0.04 | 15.42 ± 0.06 | 15.81 ± 0.04 | KAIT |
| 57701.3 | 17.69 ± 0.18 | 16.40 ± 0.06 | 15.87 ± 0.04 | 15.44 ± 0.04 | 15.92 ± 0.05 | KAIT |
| 57702.3 | 17.51 ± 0.06 | 16.36 ± 0.02 | 15.81 ± 0.02 | ... | ... | Nickel |
| 57703.3 | 17.55 ± 0.36 | 16.49 ± 0.14 | 15.91 ± 0.08 | 15.47 ± 0.08 | 15.92 ± 0.04 | KAIT |
| 57706.3 | 17.83 ± 0.46 | 16.53 ± 0.14 | 15.91 ± 0.08 | 15.58 ± 0.08 | 15.97 ± 0.05 | KAIT |
| 57707.2 | ... | ... | ... | ... | 15.93 ± 0.07 | KAIT |
| 57710.3 | 17.80 ± 0.20 | 16.55 ± 0.06 | 16.06 ± 0.04 | 15.62 ± 0.04 | 16.07 ± 0.06 | KAIT |
| 57710.3 | 17.57 ± 0.12 | 16.50 ± 0.04 | 16.00 ± 0.02 | 15.60 ± 0.04 | ... | Nickel |
| 57744.2 | 18.00 ± 0.04 | 17.00 ± 0.02 | 16.56 ± 0.02 | 16.04 ± 0.02 | ... | Nickel |
| 57753.1 | 18.09 ± 0.04 | 17.38 ± 0.06 | 16.80 ± 0.02 | 16.32 ± 0.02 | ... | Nickel |

**Extended Data Table 4:** H II Region Line Fluxes

| Line | Flux [$10^{-15}$ erg cm$^{-2}$ s$^{-1}$] |
|---|---|
| H$\beta$ | 19.2 ± 0.4 |
| [O III] $\lambda$ 4959 | 2.71 ± 0.31 |
| [O III] $\lambda$ 5007 | 9.64 ± 0.29 |
| [N II] $\lambda$ 6548 | 5.74 ± 0.05 |
| H$\alpha$ | 55.2 ± 0.3 |
| [N II] $\lambda$ 6584 | 19.4 ± 0.1 |
| [S II] $\lambda$ 6717 | 4.06 ± 0.05 |
| [S II] $\lambda$ 6731 | 2.75 ± 0.07 |
| [S III] $\lambda$ 9069 | 1.67 ± 0.02 |
| [S III] $\lambda$ 9532 | 2.62 ± 0.03 |

**Extended Data Figure 1**

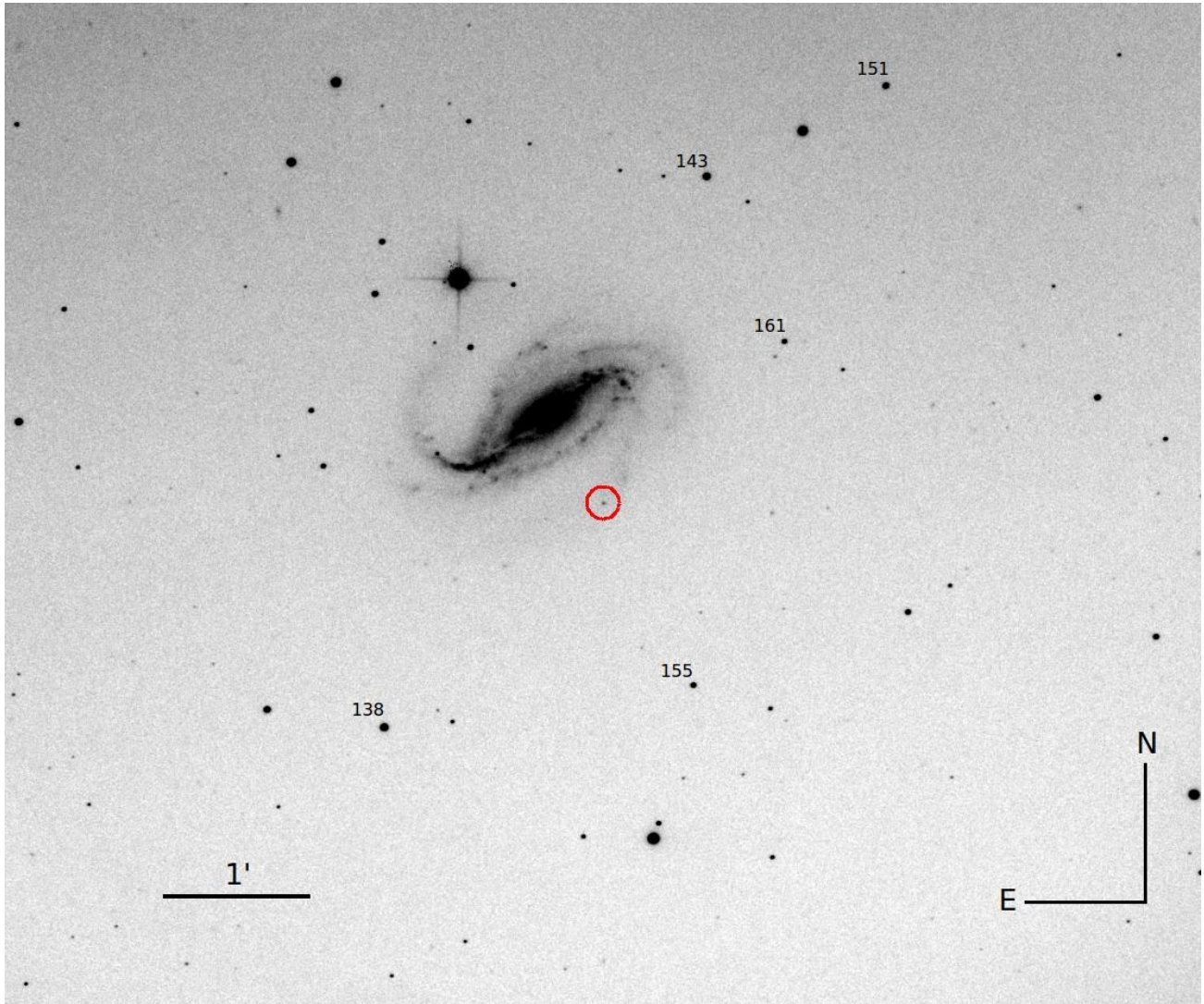

**Buso's image of SN 2016gkg in NGC 613.** The image is a combination of the final 21-image series obtained at discovery. We show only the relevant region containing the SN (red circle), its host, and the comparison stars for photometry (indicated with numbers on the upper left of each star; see Extended Data Table 1).

**Extended Data Figure 2**

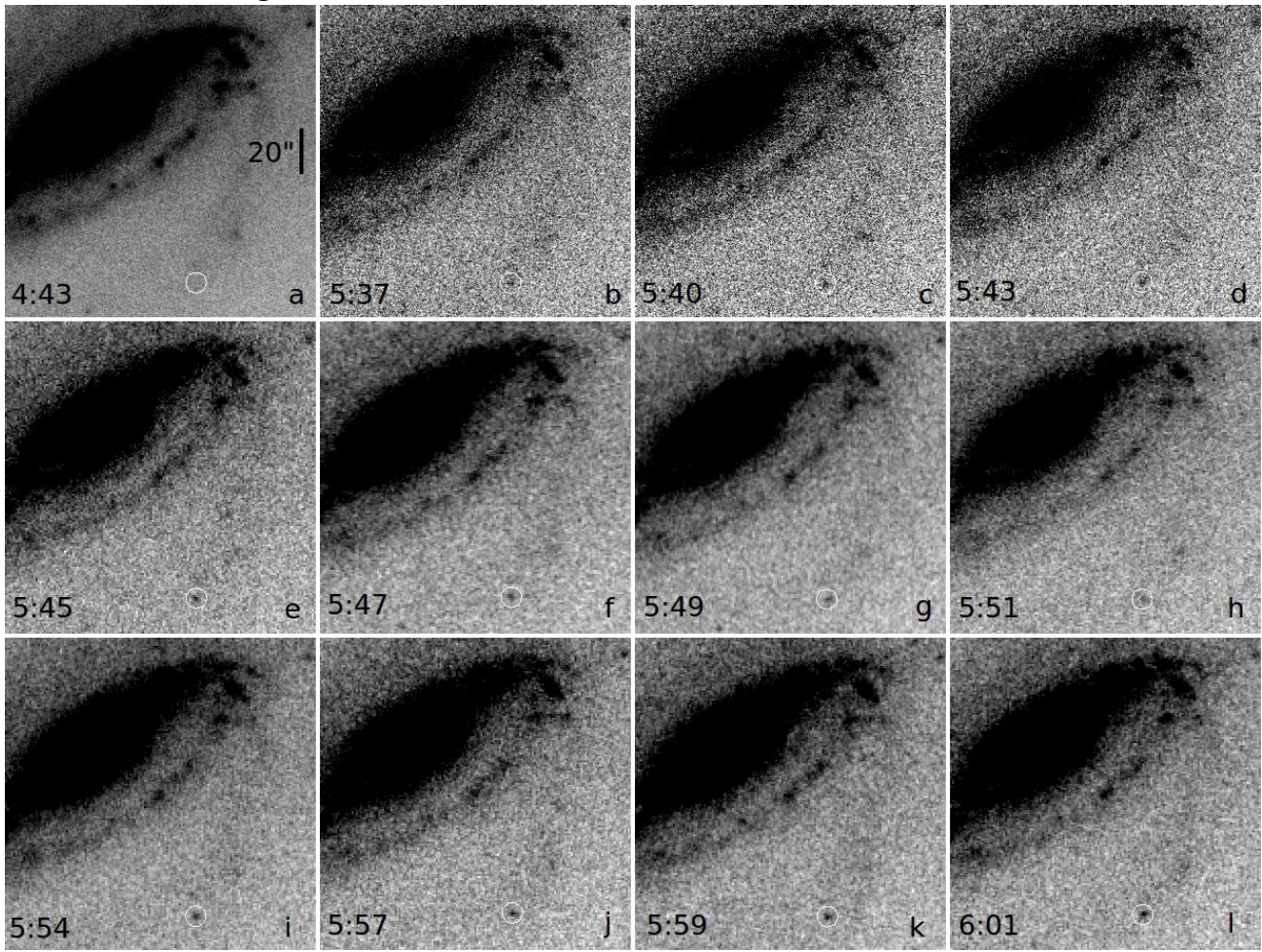

**Series of discovery images of SN 2016gkg obtained by V. Buso.** The SN location is indicated in all panels with a white circle. North is up and east is to the left. The bar in panel (a) indicates a scale of 20". Panel (a) shows a combination of 40 exposures obtained prior to the SN detection. Panels (b) through (l) display the sequence of images obtained during the initial rise as combinations of five or six individual exposures. Labels on the lower left of each panel indicate the mean UT time of the images. Photometry from the latter set of images is shown with blue diamonds in Figure 1 of the main text.

**Extended Data Figure 3**

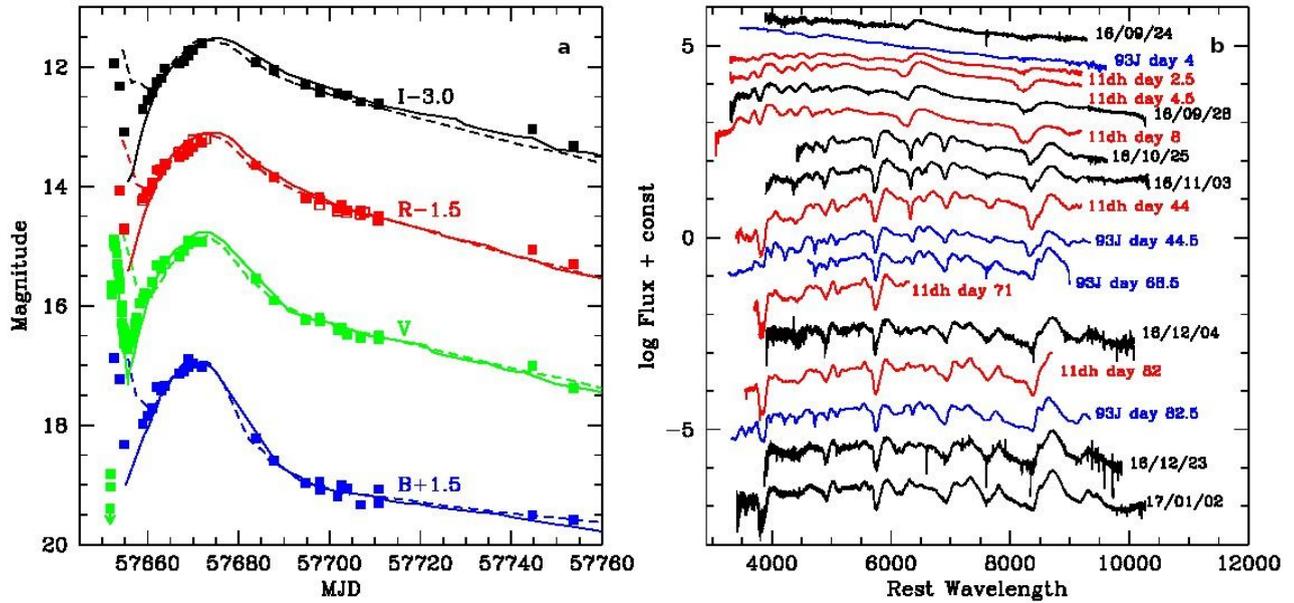

**Follow-up observations of SN 2016gkg compared with those of other Type IIb SNe.**
Panel a: *BVRI* light curves SN 2016gkg (symbols) obtained with the KAIT and Nickel telescopes. Also included are *V*-band data from V. Buso J. L. Sánchez, converted from the "Clear" band, and data from Atlas, ASAS, Swift, and LCOGT [17]. Open symbols are unfiltered data from KAIT, transformed to the *R* band. Data of Type IIb SNe 1993J [dashed lines; 74] and 2011dh [solid lines; 39, 43, 75] are included for comparison. Panel b: Optical spectra of SN 2016gkg (in black) compared with data of the Type IIb SNe 1993J [in blue; 41, 42] and 2011dh [in red; 43] at similar epochs.

**Extended Data Figure 4**

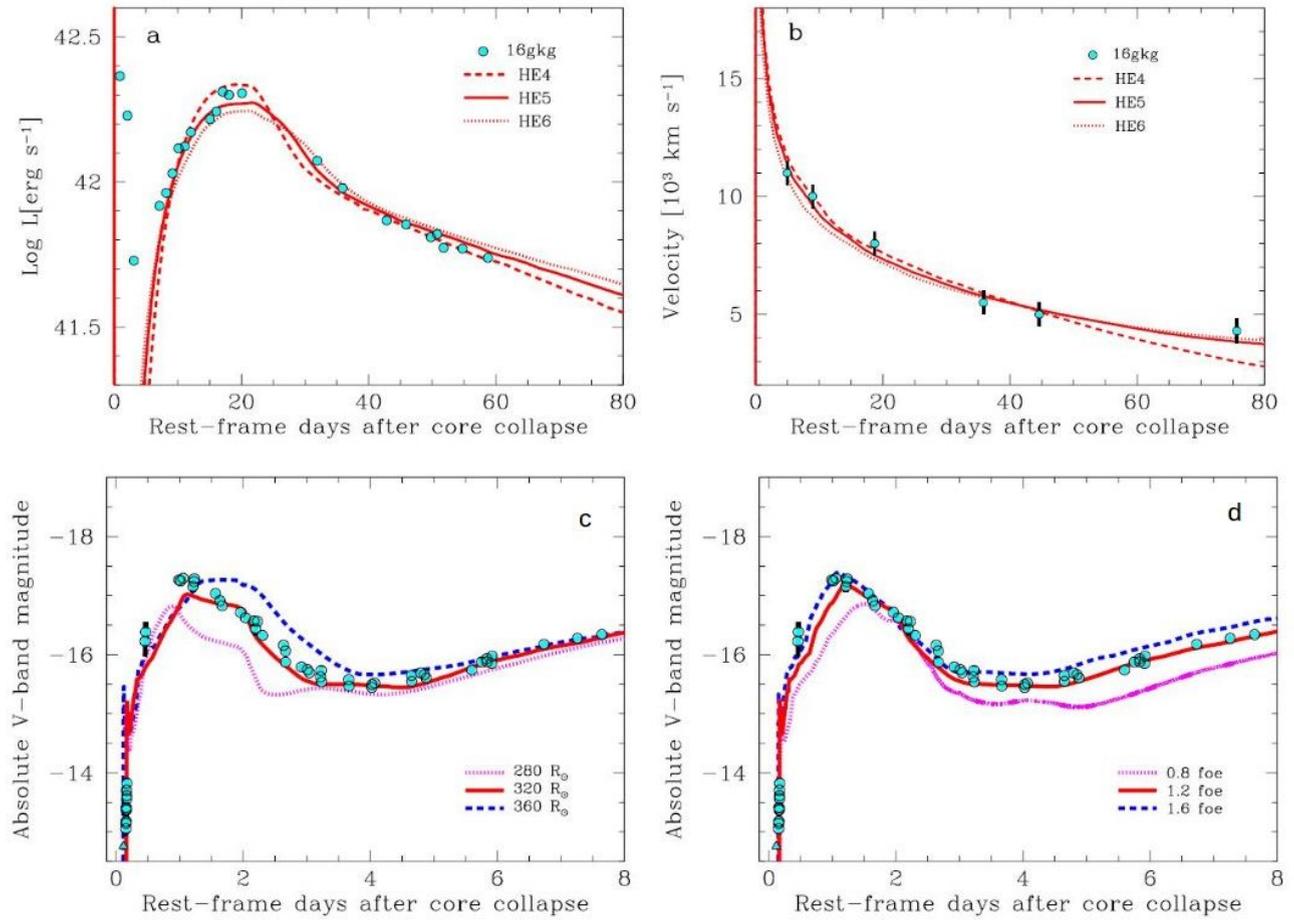

**Hydrodynamical modelling of SN 2016gkg.** Top row: model (lines) bolometric light curve (panel a) and photospheric velocity evolution (panel b) compared with observations (dots) during the $^{56}$Ni-dominated phase. No attempt was done to reproduce the initial light-curve decline (at t < 4 d). Bottom row: absolute *V*-band light curve models (lines) compared with observations (dots) during the SBO and post-shock cooling phases, for different progenitor radii (panel c) and explosion energies (panel d). Uncertainties are given as 1σ standard deviations and are shown only when they are larger than the data points.

**Extended Data Figure 5**

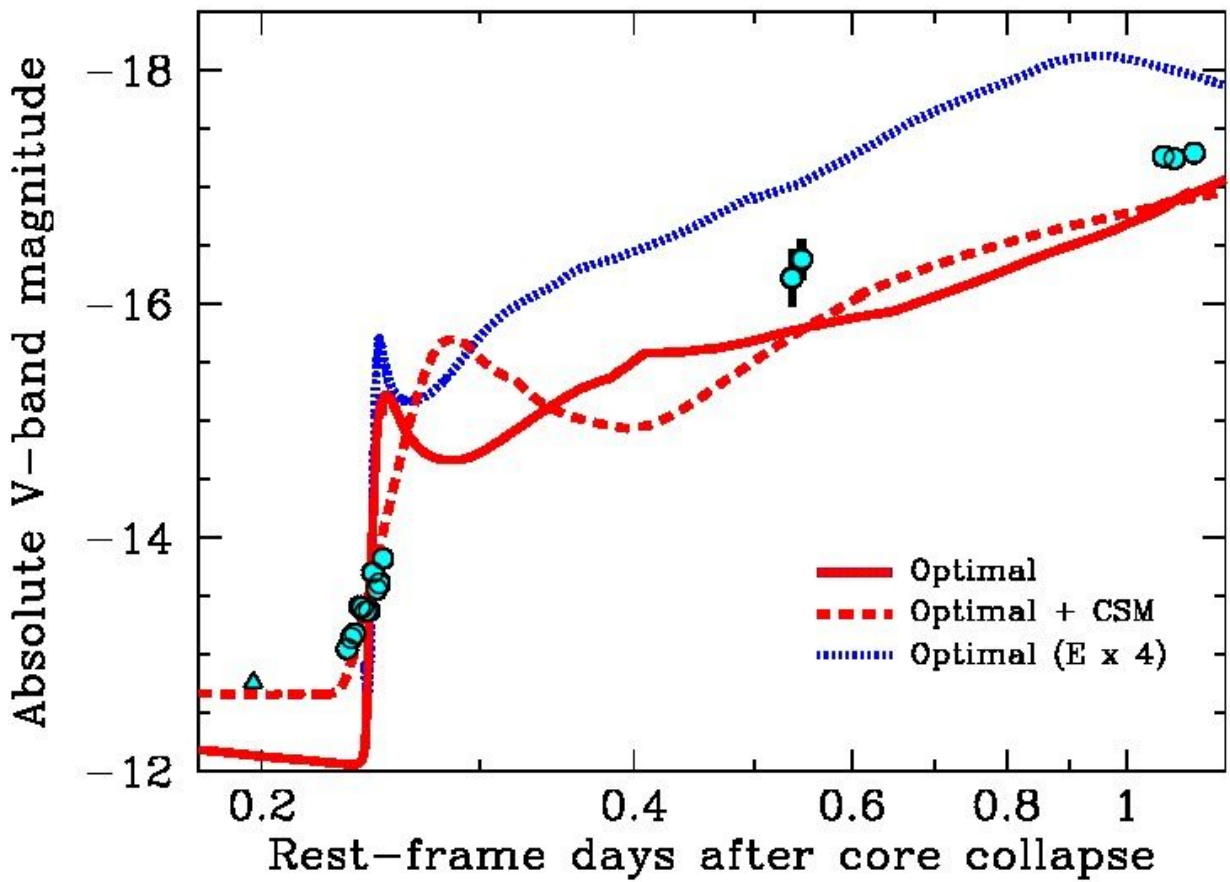

**Modelling of the initial rise of SN 2016gkg.** Absolute *V*-band of our preferred model (solid line), a similar model including some circumstellar material (CSM; dashed line) and a model with ~4 times larger explosion energy (dotted line), compared with the early-time observations (dots). The CSM is not necessarily in hydrostatic and thermal equilibrium. The presence of the CSM material reduces the slope during the SBO phase, making it even more compatible with the observations, without affecting the evolution at times later than ~1 day. Even assuming an extreme explosion energy, the resulting CP slope is substantially smaller than that during the SBO. Uncertainties are given as 1σ standard deviations and are shown only when they are larger than the data points.

**Extended Data Figure 6**

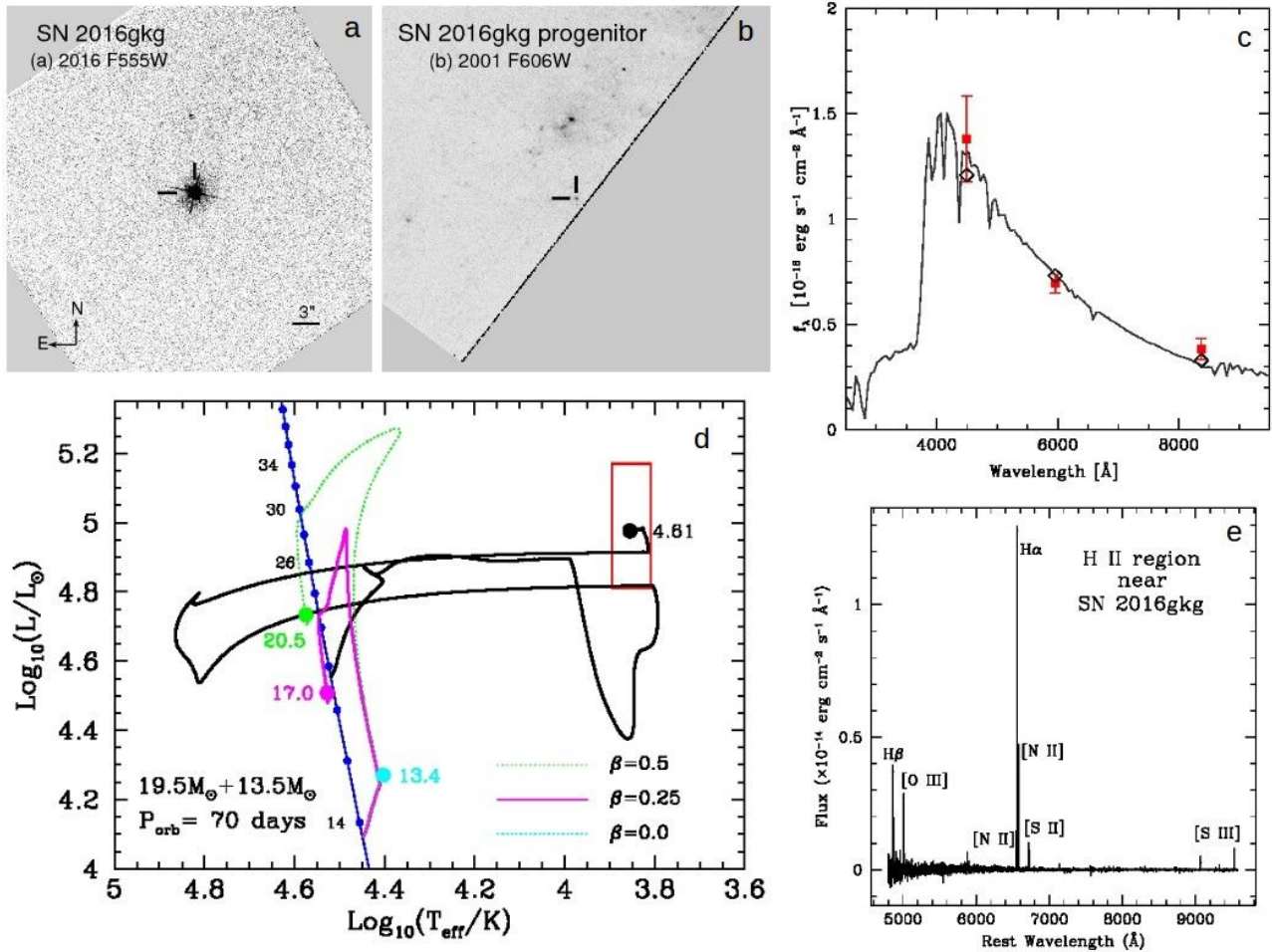

**The progenitor candidate and environment of SN 2016gkg.** (a) The HST WFC3/UVIS F555W image mosaic from 2016 October 10. (b) A portion of the HST WFPC2 F606W image mosaic from 2001 August 21. The progenitor candidate position is indicated by tick marks. (c) Stellar atmosphere SED fit (line) to the candidate HST photometry (red squares). An H II region of which we obtained a Keck DEIMOS spectrum is seen ~8".6 north of the progenitor. Uncertainties are given as 1σ standard deviations. (d) Evolutionary tracks on the HRD of our progenitor binary model (primary star in black; secondary star in cyan, magenta and green for different accretion efficiencies). Large dots indicate the endpoints of both stars, with final masses labelled, the red square shows the progenitor candidate location, and the blue line is the zero-age main-sequence with masses indicated. (e) Spectrum of a bright H II region seen in panel b, 8".6 north of SN 2016gkg.